\documentclass[prb, showpacs, preprintnumbers]{revtex4}

\usepackage {graphicx}

\begin{document}

\preprint{IGCAR/MSD/2003}
\title{\bf Systematics in the superconducting and normal state properties
 in chemically substituted MgB$_{2}$} 

\author{S. Jemima Balaselvi, A.Bharathi$^{*}$, V.Sankara Sastry, G.L.N.Reddy and Y.Hariharan} 
\affiliation{ Materials Science Division, Indira Gandhi Center
for Atomic Research, Kalpakkam, Tamil Nadu 603 102, INDIA} 
\email[E-mail: ]{bharathi@igcar.ernet.in} 
\date{ February 24, 2003} 
\begin{abstract} 
The superconducting transition temperature, T$_{C}$, the residual 
resistivity $\rho_{0}$ and
the slope of resistivity curve at high temperature, d$\rho$/dT, have been measured in a 
series of MgB$_{2}$ samples that have been chemically substituted to varying degree 
with Li or Cu at the Mg-site and by Li or Cu at the Mg-site along with C 
substitution at the B-site. DC resistivity and ac susceptibility measurements 
were employed to extract the above parameters. T$_{C}$ versus the electron count 
(estimated from simple chemical valence count arguments) shows a 
universal behaviour, with T$_{C}$ being constant at the MgB$_{2}$ value for electron 
counts lower than in MgB$_{2}$ but rapidly decreasing for larger electron counts.  The 
temperature dependence of resistivity in the normal state fits to the Bloch- 
Gruneisen formula, from which the Debye temperature,
$\theta_{D}$, and the $\rho_{0}$ are  
extracted. $\theta_{D}$ variation with T$_{C}$ is not
systematic, whereas $\rho_{0}$ versus T$_{C}$ shows a  
systematic variation that depends  on the type of the chemical substituent.
This dependence has a  
signature of the nature of the intraband/interband scattering affected by the 
chemical substitutions. d$\rho$/dT increases with C substitution, but decreases with 
Li and Cu substitution, implying that C substitution leads to
the domination of 
conductivity by the $\sigma$ band, while in the Li/Cu 
substituted samples the $\pi$ band 
dominates conduction. 
\end{abstract} 
 \maketitle
\section{1. INTRODUCTION} 
Starting with the discovery of superconductivity at 39K in 
MgB$_{2}$ \cite{nagamatsu}, there has been 
hectic activity both theoretical and experimental to unravel the origin of 
superconductivity in this system. The reduced isotope effect\cite{budko,hinks} and the 
negative pressure coefficient\cite{lorentz,tomita} of T$_{C}$ seems to indicate that MgB$_{2}$ falls in the 
category of conventional electron-phonon coupled superconductors with a large 
electron phonon coupling constant\cite{budko,hinks}. Detailed band structure calculations 
performed on the system suggest that the Fermi surface consists of two  
cylinders arising from hole-like $\sigma$ bonding bands and one 
electron like and 
another hole like 3-dimensional tubular network arising from the bonding and the 
antibonding $\pi$ bands.\cite{kortus,an} The phonon density 
of states has also been calculated 
for the system from which it is now clear that the E$_{2g}$ phonon couples non- 
linearly with holes in the $\sigma$ band and that it is also 
anharmonic.\cite{yildrim,kong} The 
presence of both $\sigma$ and $\pi$ bands at the Fermi surface 
and their different 
couplings with the phonons result in a k-dependent 
superconducting gap\cite{liu} 
hitherto not observed in earlier superconductors. An effective two gap 
superconductivity seems sufficient to describe the anomalous specific heat and 
tunneling data in MgB$_{2}$.\cite{bouquet,giubileo} Evidence 
for multigap superconductivity is now 
accruing from scanning tunneling microscopy , Raman 
scattering and  point contact
spectroscopy.\cite{iavorone,schmidt,chen}

Various chemical substitutions have been carried out primarily to 
increase T$_{C}$ and to verify several of the early theoretical predictions. 
The substituents at Mg site that have been examined are 
Al,\cite{slusky,postorino,lorentz1} Si,\cite{cimberele} alkali 
metals,\cite{zhao,syli,toulemonde} 3d transition 
metals\cite{moritomo,kazakov,xu} and 4d transition 
metals.\cite{gasporov,kalavathi} These 
substitutions have almost always resulted in a decrease in T$_{C}$, irrespective of 
whether the substituent is an electron dopant or a hole dopant, with the 
exception of Zn substitution in which an increase in T$_{c}$ of
$\sim$0.2K was observed. This  
increase was found to correlate with an expansion of the 
lattice.\cite{moritomo,kazakov} A similar 
correlation of the volume expansion with a T$_{C}$ increase was 
also observed in our 
previous study on 4d- transition metal substitution in MgB$_{2}$ where a small 
($\sim$0.5K) increase in T$_{C}$ was observed for 5\% Nb 
substitution in MgB$_{2}$.\cite{kalavathi} One of the 
systems studied in detail has been 
Mg$_{1-x}$Al$_{x}$B$_{2}$\cite{slusky} in which T$_{C}$ 
shows a monotonic decrease with substitution in the x=0.0 to 
x=0.3 composition range, there is an 
abrupt decrease in T$_{C}$ at x=0.33, beyond which again a
monotonic decrease in T$_{C}$ is  
observed\cite{postorino}. There is also an associated compression along the c-axis and along 
the a-axis resulting in a net decrease in the cell volume. Thermopower studies 
in this series showed that the charge carriers are holes and the decrease in T$_{C}$ 
correlates with the decrease in the hole density of states at
the Fermi level\cite{lorentz1} 
as a result of electron doping in the system. Band structure calculations 
carried out in  Al substituted MgB$_{2}$ indicates a decrease in the area of 
the cylindrical part of the Fermi surface with substitution, which also 
correlates rather well with the decrease in T$_{C}$ with Al
substitution.\cite{pena} Be  
substitution in MgB$_{2-x}$Be$_{x}$, results in a decrease in a- lattice parameter 
and in an increase in c- lattice parameter  resulting in a net 
increase in volume, and phase stability in this system was observed upto x=0.6. 
Thermopower measurements showed an increase in hole concentration with Be 
substitution. Thus in the Be substituted samples despite an 
increase in cell volume and an increase in the hole concentration, T$_{C}$ decreases. 
It is reasoned that the decrease in T$_{C}$ is correlated with
the decrease in 'a' lattice 
parameter which leads to a depletion of charge at the B site, causing an 
increase in phonon frequency and decrease in electron phonon 
coupling.\cite{ahn1,ahn2} 
Carbon substitution at B 
site,\cite{zhang,paranthaman,jsahn,mickelson,takenobou,bharathi} 
reported by different groups showed a 
decrease in T$_{C}$ and cell volume with increase in C content.
The decrease in T$_{C}$ is attributed to a  
decrease in hole density of states at the Fermi level due to electron doping. 
The small differences in the extent of decrease in T$_{C}$ among the various reports 
arise on account of the differences  of C solubility 
into the MgB$_{2}$ 
lattice. This variation in solubility has been attributed to the form of C 
employed in the synthesis and the method of sample preparation. We have 
reported\cite{bharathi} a C solubility upto x=0.30 in MgB$_{2-x}$C$_{x}$ and 
a decrease in T$_{C}$ of upto 
$\sim$26K in these samples, made using a home built 50 bar 
pressure lock-in set-up. 
The decrease in T$_{C}$ with C concentration matches with that observed for Al 
substitution, possibly indicating that the additional electrons due to C/Al 
substitution fill the MgB$_{2}$ bands in a rigid band manner. 
From the substitution studies, discussed above, it is clear that electron doping results in a 
decrease in T$_{C}$, and an increase in T$_{C}$ due to hole 
doping has not been observed. 

Calculations\cite{mehl} predicted an increase in T$_{C}$ by complete 
substitution of Mg by Cu and partial C substitution of B in the 
MgB$_{2}$ lattice. 
The rationale behind the prediction was that C substitution for B would result 
in an increase in stiffness of B-C bond and hence the electron-phonon coupling 
strength. Since C substitution results in electron doping, known 
to deplete T$_{C}$, 
it was thought that substitution of Cu for Mg would provide the compensating 
holes in the system. Search along similar lines led to the prediction of hole 
doped LiBC to be a high temperature superconductor,\cite{rosner} 
in which the presence of 
the B-C network results in a large electron-phonon interaction.
\noindent 
\vskip 3mm 
Early band structure calculations\cite{an} also point out that
the hole DOS in MgB$_{2}$ is  
2-dimensional in character and that an increase in the hole concentration may 
not result in an increase in T$_{C}$. It was shown that hole DOS 
is flat below E$_{F}$, 
i.e:, will remain constant with hole doping but falls off slowly with a small 
increase in electron doping and rather precipitously beyond 0.2 electron 
addition per formula unit. In our earlier study on C doped 
MgB$_{2}$, we indeed did see a 
large decrease in T$_{C}$ with large electron dopings, which
could be attributed to a precipitous decrease in the hole DOS.
In  an attempt to check the hole DOS picture further, we 
started out on the synthesis of samples with differing hole concentration 
levels by suitable Li and Cu substitution and have examined the 
variation in T$_{C}$ 
in them with electron addition by C substitution. The two initial hole dopings 
we started out with were 20\% Li substituted and 5\% Cu 
substituted MgB$_{2}$. These 
compositions were arrived at based on the determination of individual 
solubilities of Cu and Li in MgB$_{2}$ by systematic chemical substitution studies of 
Cu and Li respectively.

\noindent 
\vskip 1mm 
The different series of samples that are examined in this work are, Li and Cu 
substitution at Mg site to study the effect of holes viz., 
Mg$_{1-x}$Li$_{x}$B$_{2}$ and Mg$_{1-x}$Cu$_{x}$B$_{2}$ and 
electron doping by C substitution at B site along with hole 
doping of 20\%Li and 5\%Cu at the Mg site, viz., 
Mg$_{0.80}$Li$_{0.20}$B$_{2-x}$C$_{x}$ and 
Mg$_{0.95}$Cu$_{0.05}$B$_{2-x}$C$_{x}$. The experimental 
techniques employed in this 
study are  ac susceptibility measurement for determining T$_{C}$
and resistivity  
measurement from 300K to 4.2K, to determine T$_{C}$ and normal
state transport. The superconducting transition temperatures are 
extracted from the onset of the diamagnetic signal and from the
zero resistance.  
The temperature dependent normal state resistivity is fitted to the Bloch- 
Gruneisen formula to extract the Debye temperature $\theta_{D}$
and residual resistivity  
$\rho_{0}$. Using this $\theta_{D}$ and the measured T$_{C}$ in
the McMillan equation, the electron-phonon coupling $\lambda$ is extracted. 
The slope of the linear part of the resistivity curve in the 200K-300K 
temperature range has also been determined in each of the 
samples, in order to quantify the magnitude of the temperature 
dependence of resistivity. The various 
measured parameters on these samples have been compared with 
that in MgB$_{2-x}$C$_{x}$.\cite{bharathi} 
The paper is organized as follows. In section.2 the experimental details are 
mentioned. In section.3 we present the various results along with the corresponding 
discussions in different subsections. Section.4 provides the summary and 
conclusions.

\section{2. EXPERIMENTAL } 
Samples of nominal composition Mg$_{1-x}$Li$_{x}$B$_{2}$ 
[x=0.1, 0.2, 0.3], Mg$_{1-x}$Cu$_{x}$B$_{2}$ 
[x=0.01, 0.02, 0.025, 0.05], Mg$_{0.80}$Li$_{0.20}$B$_{2-x}$C$_{x}$ 
[x=0.1, 0.2] and Mg$_{0.95}$Cu$_{0.05}$B$_{2-x}$C$_{x}$ 
[x=0.025, 0.05, 0.15, 0.3] were prepared by the standard solid-vapour technique 
using Mg powder (99.9\%), amorphous boron (99\%), carbon soot
(99\%) obtained from  
fullerene synthesis, Li pieces (99\%) and Cu powder (99.9\%). The stoichiometric 
quantities are weighed, mixed and  compacted into a Ta crucible
and heat treated
at 900$^{0}$C for 1hour 30 minutes under a locked-in Ar pressure
of 50 bar.\cite{bharathi}  Li was 
loaded inside a dry box under Ar atmosphere. Weight loss was consistently 
recorded to be less than 1\% indicating that the nominal composition is 
preserved even after the heat treatment in all the samples. Samples so obtained 
were of $\sim$30\% theoretical density and suitable for resistivity measurements by 
appropriate slicing. However some samples that were not compacted
properly, resulted in porous powders, in which resistivity
contacts could not be achieved. The samples were characterized by powder X-ray diffraction 
in a STOE diffractometer using Cu-K$_{\alpha}$ radiation in the Bragg-Brentano geometry. 
Susceptibility measurements were done by tracing the diamagnetic signal using an 
ac mutual inductance technique at a measuring frequency of 941 Hz for which 25 
mg of finely powdered samples were used. The resistance measurements were done on 
$\sim$1mm thick sliced pieces in the standard four probe geometry using, 42 SWG Cu 
leads with silver paint as contact glue. The resistivity in each
sample was measured in the Van der pauw geometry  
at room temperature, using which resistivity values at all the temperatures 
could be determined. For  both susceptibility and resistivity 
measurements, the temperature variation from 300K to 4.2K was obtained by using a 
dipstick setup in the which the temperature was measured using a
Si-diode thermometer with an excitation current of 10$\mu$A and
the data was collected through a PC, interfaced by IEEE 488  
card.  

\section{\bf 3. RESULTS AND DISCUSSION} 
\noindent 
\subsection{\bf 3.1 XRD measurements} 
From the phase purity analysis of the XRD data it is clear that the Li 
solubility in MgB$_{2}$ is at least 20\% (x=0.2), while that of
Cu is only 5\% (x=0.05). Carbon  
substitutes upto a fraction of x=0.30 in a  phase pure form in the 
cation substituted samples. The lattice 
parameters 'a' and 'c' and the cell volume obtained from an
analysis of the XRD  
data using the STOE program are shown in Fig. 1, as a function of
'x',  where x is the fraction of the substituent whose
concentration is varied in that particular series.  It can be seen from the
figure that 'c' parameter  
remains more or less unchanged with substitution in all the
series studied. For  
cationic substitutions, the 'a' parameter also remains constant,
resulting in no  
change in volume for samples that are substituted only at the Mg
site. The lack  
of change in the lattice parameters with Li substitution in Fig. 1 is at 
variance with earlier studies\cite{zhao} where a substantial
decrease in 'a' parameter was  
observed with  Li substitution. From Fig. 1, it is seen that with C 
substitutions the lattice parameter along 'a' decreases monotonically with 
increasing concentration with a corresponding decrease in the
lattice volume. The extent  
of the decrease is the largest for the MgB$_{2-x}$C$_{x}$, intermediate in 
Mg$_{0.95}$Cu$_{0.05}$B$_{2-x}$C$_{x}$ and small in 
Mg$_{0.80}$Li$_{0.20}$B$_{2-x}$C$_{x}$. The lattice constant 
remaining unchanged in the Li substituted samples can be rationalized based on 
the fact that the ionic radii of Li$^{+}$ of Mg$^{++}$ are not
very different. The lack  
of change due to Cu substitution can be reconciled with, from the
fact that Cu is  
soluble only to 5\% and a difference in the ionic radii may not reflect as a 
measurable change in lattice constant. The decrease in the a-parameter with C 
content can result due to the smaller covalent radius of C in comparison with 
that of B. The smaller decrease in the a-parameter with C 
substitution in the Li 
and Cu substituted samples is surprising. The differences in these lattice 
parameter variations suggest that the electron concentration present in the 
sample also plays an important role in determining the equilibrium lattice 
constants.
\begin{figure}
\includegraphics*[width=7cm,height=9cm]{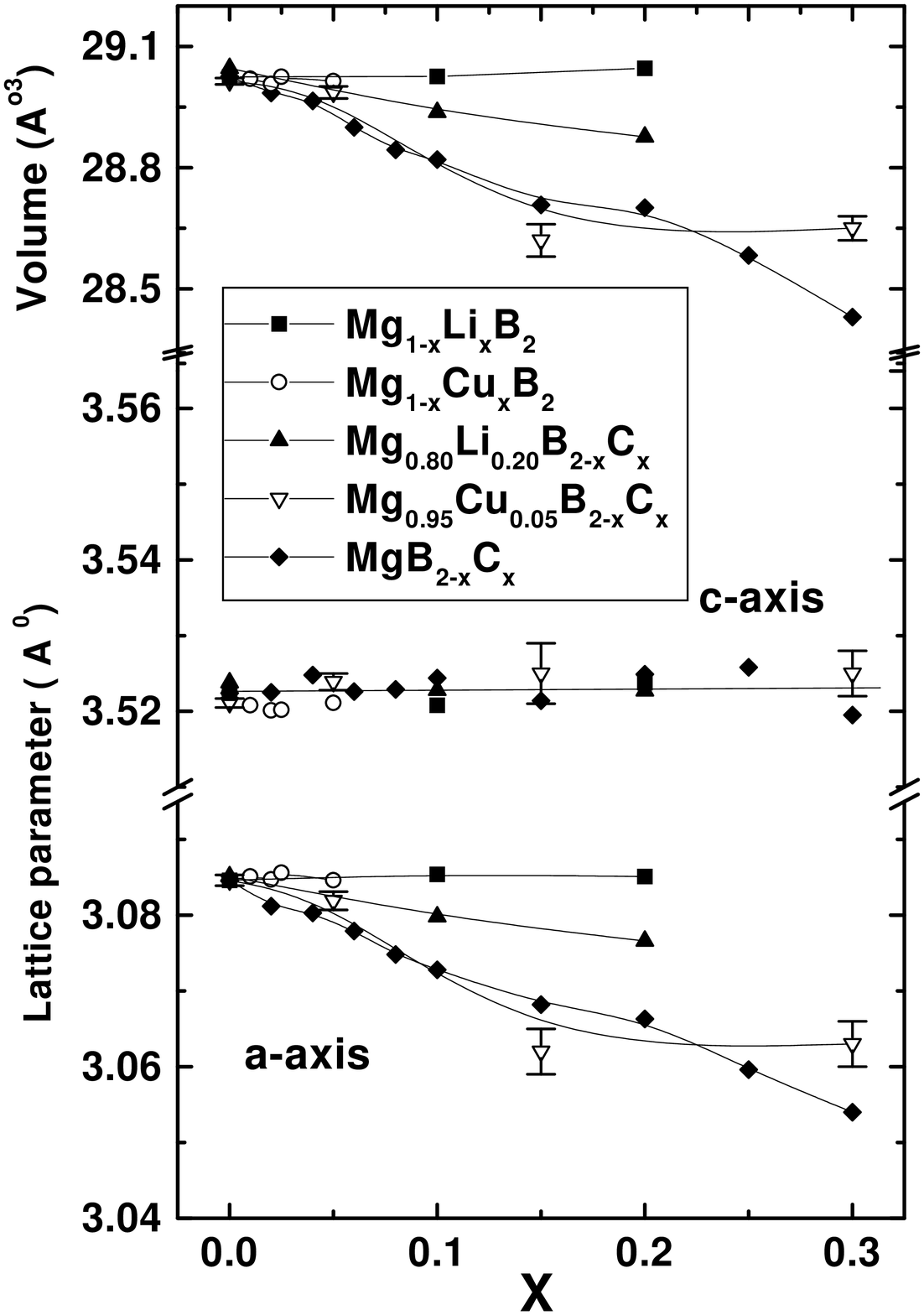}
\caption{\label{fig1} Plot of the lattice constants 'a' and 'c' and the lattice cell volume, 
obtained from XRD patterns, as a function of the varying substituent
fraction,'x'. The solid  lines are guide to the eye.}
\end{figure} 
\subsection{ 3.2 T$_{C}$ from ac susceptibility and from dc resistivity } 
\noindent 
In Fig.2 is shown the variation of the diamagnetic signal for
MgB$_{2}$ in the 4.2K to 50K temperature range. 
The T$_{C}$ of 39.6K is deduced by reading off the value of
temperature at 10\% of the total diamagnetic  signal determined
from the onset. The transition width 
($\Delta$T$_{C}$), determined from the difference in
temperatures at 90\% and 10\% of the 
total diamagnetic signal, was 1.75K. 
The variation of T$_{C}$ along each 
series is shown in the inset of Fig.2. The observed values of
T$_{C}$, $\Delta$T$_{C}$ and volume  
fraction of the net diamagnetic signal for the various samples are tabulated in 
Table.1. 
\begin{figure}
\includegraphics*[width=7cm,height=7cm]{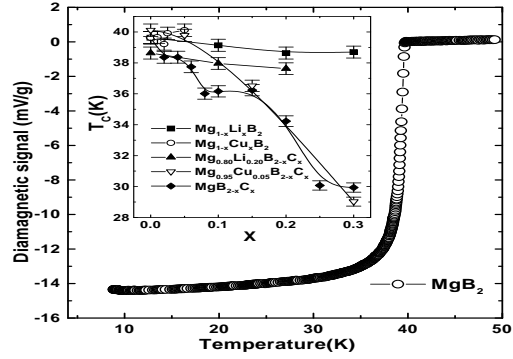}
\caption{\label{fig2} Plot of diamagnetic signal versus temperature in the
range 4.2K-50K for  pristine MgB$_{2}$. In the inset is shown
the variation in T$_{C}$ with fraction 'x' of  
varying substituent in the different series studied. The temperature at
the 10\% signal measured from the onset  
is taken as T$_{C}$.} 
\end{figure}
\begingroup
\squeezetable[width=8cm,height=8cm]
\begin{table*}
\caption{Table.1 Data from susceptibility measurements; temperature
corresponding to 10\%  
of the total diamagnetic signal, measured from the onset ie
recorded T$_{C}$; difference  
between the temperatures corresponding 10\% signal and 
90\% signal is recorded as $\Delta$T$_{C}$ and the corresponding
magnitude of the signal is the volume fraction.} 
\begin{tabular}{|l|l|l|l|l|l|l|} 
\hline
Mg$_{1-y}$M$_{y}$B$_{2-x}$C$_{x}$&M&y& x&  T$_{c}$&
$\Delta$T$_{c}$(K)& Vol.Frac.(mV/gm)\\
\hline
MgB$_{2}$&  & 0& 0& 39.6&1.75&14.0\\
MgB$_{2-x}$C$_{x}$& & 0& 0.02&38.3&2.41&12.9\\
                  & & 0& 0.04&38.4&4.13&13.6\\
                  & & 0& 0.06&37.7&4.51&13.1\\
                  & & 0& 0.08&36.0&4.11&12.8\\
                  & & 0& 0.10&36.1&8.08&13.5\\
                  & & 0& 0.15&36.2&9.17&11.3\\
                  & & 0& 0.20&34.2&14.09&11.5\\
                  & & 0& 0.25&30.1&14.66&11.9\\
                  & & 0& 0.30&29.9&15.12& 9.0\\
                  
Mg$_{1-y}$Li$_{y}$B$_{2}$&Li & 0.1& 0&39.1&4.5&7.8\\
                         & & 0.2& 0&38.8&4.9&7.6\\
                         & & 0.3& 0&38.7&5.7&4.9\\

Mg$_{0.8}$Li$_{0.2}$B$_{2-x}$C$_{x}$& & 0.2& 0.1&38.0&3.3&9.6\\
                                    & & 0.2& 0.2&37.6&5.9&10.1\\
Mg$_{1-y}$Cu$_{y}$B$_{2}$&Cu & 0.01& 0&39.7&3.2&7.1\\
                         &   & 0.02& 0&39.2&3.5&9.7\\
                         &   &0.025& 0&39.9&2.7&8.7\\
                         &   & 0.05& 0&40.11&2.8&7.7\\
Mg$_{0.95}$Cu$_{0.05}$B$_{2-x}$C$_{x}$& & 0.05& 0.05&39.8&3.4&8.5\\
                                      & & 0.05& 0.15&36.5&8.8&6.6\\
                                      & & 0.05& 0.30&29.0&9.5&4.9\\
\hline
\end{tabular}
\end{table*}

In Fig.3 is shown the temperature dependence of resistivity curve for pristine 
MgB$_{2}$. The value of RR defined as the ratio of the
resistivity at 300K to the  
resistivity at 40K ($\rho$(300K)/$\rho$(40K)), observed for 
MgB$_{2}$ was $\sim$6 with a T$_{C}$ of 39.4K 
at zero resistance, and the  transition width $\Delta$T$_{C}$
determined as the difference in the onset and downset temperature is $\sim$0.3K.
The T$_{C}$ 
variation across the different series, is shown in the inset of
Fig.3 and in Table. 2.  
The variation of T$_{C}$ as a function of substituent from the
resistivity data is in  
general agreement with that observed by susceptibility
measurements. It is clear  
from Fig.3 that in the series Mg$_{1-x}$Cu$_{x}$B$_{2}$,
T$_{C}$ remains almost   
constant, whereas in 
Mg$_{1-x}$Li$_{x}$B$_{2}$ it shows a small decrease. The transition
width is  $\sim$0.3K for all 
these concentrations. In  
contrast, in all the carbon substituted samples a systematic
decrease in T$_{C}$ is  
seen and  the extent of decrease in T$_{C}$ is dependent on the 
amount of cation substituted. Across the series
Mg$_{0.95}$Cu$_{0.05}$B$_{2-x}$C$_{x}$ a decrease in  
the T$_{C}$ of 10K for a maximum carbon substitution of x=0.3 is
observed, compared to  
a T$_{C}$ decrease of 14K in the MgB$_{2-x}$C$_{x}$ series for a
comparable  carbon  
content.  Whereas, in the  
Mg$_{0.80}$Li$_{0.20}$B$_{2-x}$C$_{x}$ series only a decrease of
$\sim$1K is observed and the transition  
width is also very small of $\sim$0.5K for all the samples in
this particular series. For the other carbon substitutions, as in susceptibility
measurements the  
transition width is seen to increase with the degree of
substitution. The 
differences in the variation of T$_{C}$ for a similar extent of carbon 
substitution, therefore hints at the dependence of T$_{C}$ on other
factors. We investigate one such possibility below.
\begin{figure}
\includegraphics*[width=7cm,height=7cm]{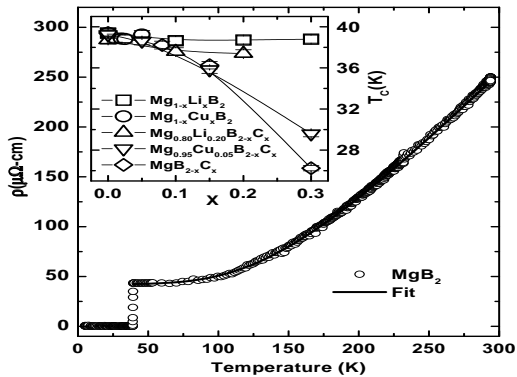}
\caption{\label{fig3} Fig. 3 Plot of the resistivity versus temperature from 4.2K-300K for pristine 
MgB$_{2}$.  The solid line shows the fitted curve resistivity curve
to the Bloch-Gruneisen formula. In the inset is shown the
variation in T$_{C}$ with 
the fraction of substituent  
x for the different series studied. Solid lines in the inset are a guide to
the eye.} 
\end{figure}
\subsection{ 3.4 T$_{C}$ versus electron count } 
We define a parameter  N$_{excess}$,
denoting the excess charge (electron/hole) with 
respect to MgB$_{2}$ as 
\begin{equation} 
\label{Eq. 1}
 N_{excess} = x - y
\end{equation} 
where, x is the fraction of the divalent anion substituent
(assumed to donate one  
electron in excess of MgB$_{2}$ for each atom substituted per
formula unit)  and y is 
the fraction of monovalent cation (which supplies one extra hole as compared to 
that in MgB$_{2}$ for one atom substituted per formula unit). The general formula for 
a representative sample studied is given by 
Mg$_{1-y}$M$_{y}$B$_{2-x}$C$_{x}$, where M=Li or Cu. 
N$_{excess}$ can be taken as a qualitative measure of the 
valence electron concentration,  with respect to that in MgB$_{2}$. N$_{excess}$ 
is zero for MgB$_{2}$, positive for electron doped MgB$_{2}$ and negative for hole doped 
MgB$_{2}$. A plot of N$_{excess}$ against T$_{C}$ is shown in Fig.4, from which it is clear 
that the T$_{C}$ remains almost constant for N$_{excess}$ $<$ 0 viz., with increase in the 
hole concentration in MgB$_{2}$, whereas it decreases with increase in electron 
concentration viz., for N$_{excess}$ $>$ 0. The prominent feature in Fig.4 is that T$_{C}$ 
variation is the same for  all the samples, albeit they belong to 
different series, depending purely on the electron count in the sample. To 
illustrate this point, focusing at N$_{excess}$=0 in Fig.4, 
obtained from MgB$_{2}$, 
Mg$_{0.80}$Li$_{0.20}$B$_{1.80}$C$_{0.20}$ and
Mg$_{0.95}$Cu$_{0.05}$B$_{1.95}$C$_{0.05}$, one sees  that the T$_{C}$s are 
nearly the same. The T$_{C}$ versus N$_{excess}$ plot in Fig.4 holds a striking 
resemblance to the hole DOS versus energy curve shown by An and 
Pickett.\cite{an} It was 
remarked in that paper that T$_{C}$ will not increase with hole doping whereas 
electron doping would result in a decrease in T$_{C}$. From Fig.4 it appears that the T$_{C}$ 
remains fixed with hole doping, whereas it shows a decrease with electron 
doping, which is smooth upto an electron doping level of $\sim$0.2, beyond which T$_{C}$ 
drops faster in agreement with the calculation.\cite{an}
\begin{figure}
\includegraphics*[width=7cm,height=7cm]{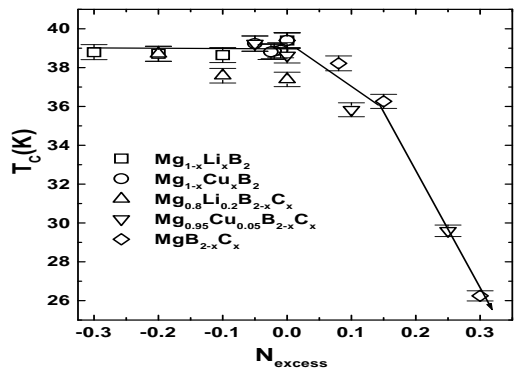}
\caption{\label{fig4}Plot of N$_{excess}$ (defined in the text) against T$_{C}$ for all the samples. The 
symbols for the different series are marked in the legend. The
solid line is a  
guide to the eye. } 
\end{figure}
\begingroup
\begin{table*}
\caption{Data from resistivity measurements; x, y and
N$_{excess}$ are described in  
the text. The temperature corresponding to zero resistivity is
shown as T$_{C}$; $\theta_{D}$,  
$\rho_{0}$, $\rho_{1}$ parameters obtained from fitting the
normal state resistivity to the  
Bloch-Gruneisen formula; electron-phonon interaction parameter,
$\lambda$ calculated  
using McMillan's equation and the slope of $\rho$(T) in the
200K-300K range.}
\begin{tabular}{|l|l|l|l|l|l|l|l|l|}
\hline

Mg$_{1-y}$M$_{y}$B$_{2-x}$C$_{x}$&N$_{excess}$&T$_{C}$&
$\theta_{D}$&$\rho_{0}$&$\rho_{1}x10^{-3}$& 
Cx10$^{3}$& $\lambda$& d$\rho$/dT\\
         & (x-y) &(K)&(K) &$\mu$$\Omega$-cm&$\mu$$\Omega$-cmK$^{-2}$&$\mu$$\Omega$-cmK& &\\
\hline

MgB$_{2}$&0  & 39.4& 911.9& 41.92&0.2536&3.6761&0.8408&1.2083\\
         &   &     &      &      &      &      &      &      \\
MgB$_{1.92}$C$_{0.08}$&0.08 & 38.2& 823.4&173.76&1.4118&2.4189&0.8784&1.3238\\
MgB$_{1.70}$C$_{0.30}$&0.30 & 26.2& 528.8&948.2&3.339&1.2801&0.9123&2.2609\\
         &   &     &      &      &      &      &      &      \\
Mg$_{0.90}$Li$_{0.10}$B$_{2}$&-0.10&38.6&992.7&30.079&0.5079&1.9901&0.8015&0.7683\\
Mg$_{0.80}$Li$_{0.20}$B$_{2}$&-0.20&38.7&920.1&22.651&0.2011&0.9223&0.8335&0.3779\\
         &   &     &      &      &      &      &      &      \\
Mg$_{0.80}$Li$_{0.20}$B$_{1.90}$C$_{0.10}$&-0.10&37.6&851.9&99.569&0.2705&1.5426&0.8545&
0.6343\\
Mg$_{0.80}$Li$_{0.20}$B$_{1.80}$C$_{0.20}$&0&37.4&674.4&271.69&1.3174&2.6158&0.9755&
1.6238\\
         &   &     &      &      &      &      &      &      \\
Mg$_{0.99}$Cu$_{0.01}$B$_{2}$&-0.01&38.9&936.5&20.550&0.2587&1.1867&0.8280&0.4510\\
Mg$_{0.98}$Cu$_{0.02}$B$_{2}$&-0.02&38.8&963.4&34.245&0.4864&1.8878&0.7594&0.7715\\
Mg$_{0.975}$Cu$_{0.025}$B$_{2}$&-0.025&38.8&928.7&16.971&0.2625&1.0823&0.8305&0.4480\\
Mg$_{0.95}$Cu$_{0.05}$B$_{2}$&-0.05&39.2&852.6&27.226&0.2506&1.4668&0.8780&0.6276\\
         &   &     &      &      &      &      &      &      \\
Mg$_{0.95}$Cu$_{0.05}$B$_{1.95}$C$_{0.05}$&0&38.6&800.1&43.603&0.3159&1.3653&0.8979&0.6455\\
Mg$_{0.95}$Cu$_{0.05}$B$_{1.85}$C$_{0.15}$&0.10&35.8&785.9&137.64&0.2789&1.1841&0.8698&
0.5686\\
Mg$_{0.95}$Cu$_{0.05}$B$_{1.70}$C$_{0.30}$&0.25&29.6&851.9&235.43&0.4922&0.8002&0.7594&
0.4978\\
\hline
\end{tabular}
\end{table*}
 
\subsection{3.5  Analysis of normal state resistivity} 
The temperature dependence of the normal state resistivity in
MgB$_{2}$,  suggests 
the dominance of phonon scattering.\cite{finnemore,canfield,putti} 
The normal state 
resistivity  for each of the data from the different series
investigated  could be fitted to 
the Bloch-Gruneisen 
formula, in the 40K and 300K range\cite{putti} using 
\begin{equation}
\label{Eq. 2}
\rho(T) = \rho_{0} + \rho_{1}T^{2} + C \rho_{ph}(T)
\end{equation}
where $\rho_{ph}$(T) is given by
\begin{equation}
\label{Eq. 3}
\rho_{ph}(T) = (m-1)\theta_{D}(\frac{T} {\theta_{D}})^{m}
\int_{0}^{\frac{\theta_{D}} {T}} dZ \frac {Z^{m}} {(1-e^{-z})(e^{z}-1)}
\end{equation} 
$\rho_{0}$ is the impurity scattering contribution to 
resistivity, $\rho_{1}$ denotes 
the magnitude of the electron-electron interaction parameter, 
m=5, $\theta_{D}$ is the Debye temperature and C is a constant.
The fit obtained for MgB$_{2}$,  
is shown in Fig.3 along with the experimental data, from which it is apparent 
that the fit is excellent. A similar quality of fit has been
obtained for each of 
the $\rho$(T) data in all the series. From the fit, the value of 
$\rho_{0}$, $\rho_{1}$, C and $\theta_{D}$ 
were extracted, which are shown in Table 2. The small values of
$\rho_{1}$ suggest that  the
contribution from electron-electron scattering to the transport
is negligible in  
the system. The $\rho_{0}$ obtained from the Bloch-Gruneisen
fits are in close agreement with $\rho$(40K), the measured
resistivity prior to the superconducting transition. It can be
seen from Table. 2 that
 $\theta_{D}$ decreases with C substitution, but the
extent of the observed  
decrease is very large in comparison to that expected from mass
considerations  
alone. In the case of Li and Cu substitutions viz., 
Mg$_{1-x}$M$_{x}$B$_{2}$ (M=Li,Cu), $\theta_{D}$ is seen to 
remain constant, 
even though a large increase in $\theta_{D}$ in the former and 
decrease in the latter is expected. Carbon 
substitution on Li substituted samples also 
shows a similar decrease in $\theta_{D}$ as in 
MgB$_{2-x}$C$_{x}$. But with Carbon substitution, in the 5\% Cu 
substituted samples $\theta_{D}$ remains constant in contrast 
to the decrease observed 
in MgB$_{2-x}$C$_{x}$ and Mg$_{0.80}$Li$_{0.20}$B$_{2-x}$C$_{x}$ 
samples. These $\theta_{D}$ variations versus the 
observed T$_{C}$ are shown in Fig.5, from which it is apparent 
that a systematics does not emerge. It should however be mentioned 
that the lack of systematics could have its origin on the fact 
that the Bloch-Gruneisen fits have been made assuming that the 
conductivity arises from a single dominant band; which is
known to be true for pristine MgB$_{2}$. The good fits to this
formula  would imply  the dominance of single band in the
substituted samples also, excepting for the 
fact that  the dominant band could be different as will
become apparent from the sections to follow. 
\noindent 
\vskip 1mm 
From the McMillan equation, 
\begin{equation}
\label{Eq. 4} 
T_{c}=(\frac{\theta_{D}}{1.45}) exp[-\frac{1.04(1+\lambda)}{(\lambda-\mu^{*}(1+0.62\lambda)}] 
\end{equation} 
using the measured T$_{C}$ and the extracted $\theta_{D}$, the electron 
phonon coupling constant, $\lambda$ is 
computed with $\mu^{*}$=0.15. The calculated values (cf. Table 2) for the entire series 
falls in the range of 0.7 and 1.0, which is in general agreement with values 
reported from theoretical calculations and specific heat measurements. 
\begin{figure}
\includegraphics*[width=8cm,height=8cm]{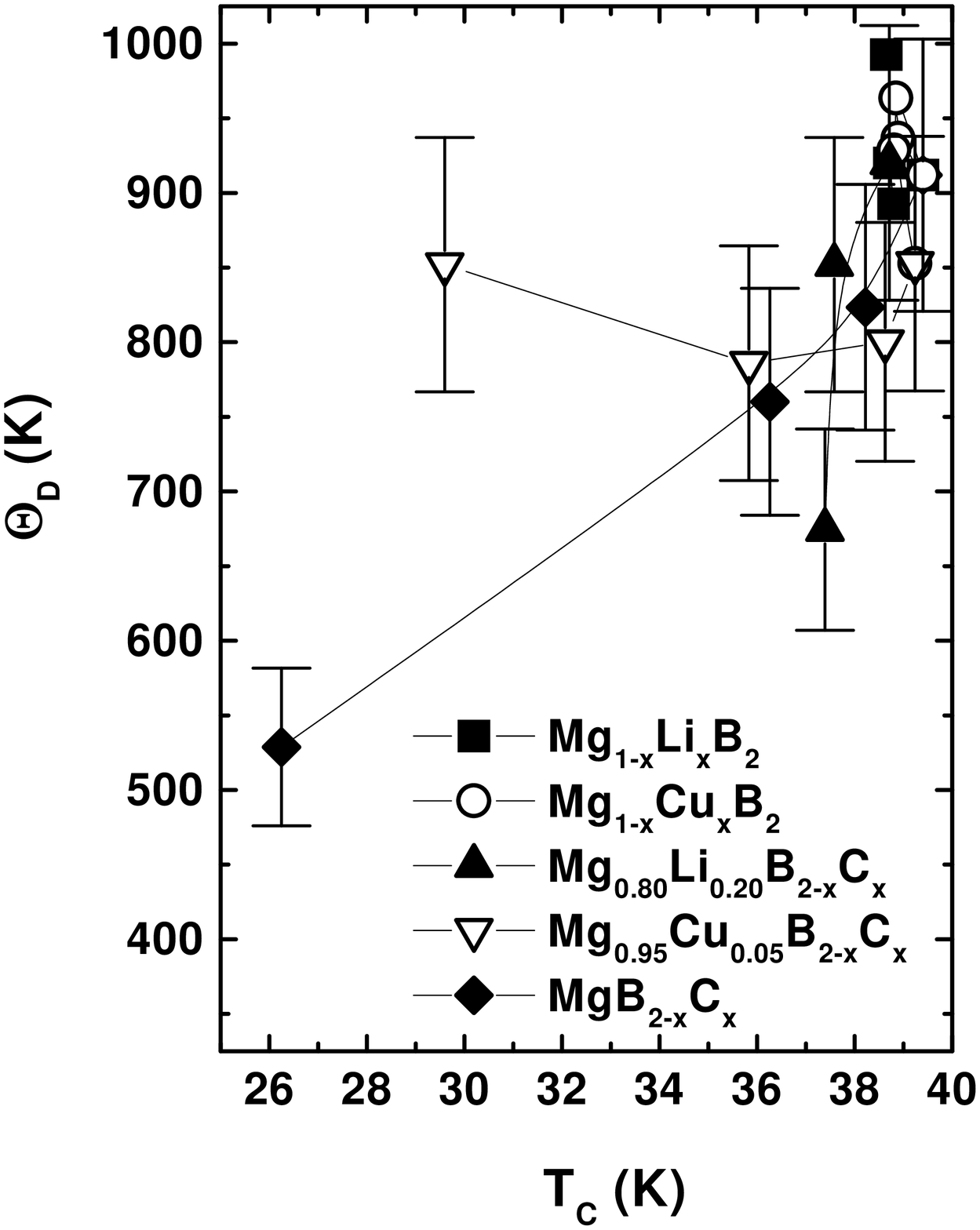}
\caption{\label{fig5}Plot of T$_{C}$ against $\theta_{D}$ obtained from the Bloch-Gruneisen fit for all 
samples studied. Solid lines are guides to the eye. } 
\end{figure}

\subsection{3.6 $\rho_{0}$ versus T$_{C}$ correlation} 
In the MgB$_{2}$ system Matthiesen's rule is violated in that
samples that have a large residual resistivity also show a
stronger temperature dependence of resistivity.\cite{mazin2} 
Further, despite the fact that impurity scattering is  detrimental to
superconductivity in a  
multiband system,\cite{golubov} T$_{C}$ is robust to the residual resistivity  
variations in MgB$_{2}$. These issues have been discussed in
a recent calculation\cite{mazin}, which suggest that the absence of interband 
scattering between the $\sigma$ 
and $\pi$ bands in MgB$_{2}$ and the dominance intraband
$\sigma$-$\sigma$ and $\pi$-$\pi$ scattering is primarily  
responsible for this unusual behaviour. It has been shown  that
the unique
electronic structure  
of MgB$_{2}$ makes $\sigma$-$\pi$ hybridization unfavourable, 
an important pre-requisite for 
the occurrence of an interband scattering event. Further it has
also been demonstrated 
that by introduction of scattering sites in the Mg sublattice, 
$\sigma$-$\pi$ hybridization is not 
significantly altered. Calculating the T$_{C}$ for different
intraband/interband scattering ratios, it has been shown 
that the slope of the plot of $\rho_{0}$ versus T$_{C}$ which is 
negligible for large 
intraband scattering cross section increases progressively with increase in 
interband scattering. These results have been compared with the
experimental T$_{C}$ values that have been obtained for pristine
MgB$_{2}$ from different laboratories.
\noindent 
\vskip 1mm 

Since in our present study the T$_{C}$ and the resistivity behaviour has been 
examined for a variety of chemical substitutions at the Mg-site, B-site and 
both Mg and B sites, it appears appropriate to see if any systematics in the 
variation of $\rho_{0}$ with T$_{C}$ can be discerned from our data. In Fig.6
is shown the plot of $\rho_{0}$  
versus T$_{C}$ for the different series examined in this work. 
From Fig.6 it is clear that there is a systematic variation of 
T$_{C}$ with $\rho_{0}$ 
within each series. In the inset of Fig.6 is shown the variation
of T$_{C}$ with $\rho_{0}$  
by monovalent cation substitution, which appears almost flat.
Taking the cue from calculations\cite{mazin}, these results
suggest that the electronic conduction in the cation substituted
samples  is dominated by intraband scattering.\cite{mazin} 
In the C substituted series the variation of T$_{C}$ 
with $\rho_{0}$ is much 
larger for similar extents of  substitution. For example in 
Mg$_{0.95}$Cu$_{0.05}$B$_{2-x}$C$_{x}$, 
the fall of T$_{C}$ with $\rho_{0}$ is largest, followed by 
that in MgB$_{2-x}$C$_{x}$, and  it 
is smallest for the Li substituted system viz., 
Mg$_{0.80}$Li$_{0.20}$B$_{2-x}$C$_{x}$. Comparing this 
with the calculated variation of T$_{C}$ with $\rho_{0}$, 
\cite{mazin} clearly suggests that the 
interband scattering starts playing a role in determining $\rho_{0}$ in the 
carbon substituted samples. This increase in interband scattering could arise 
from the presence of an excess p$_{z}$ electron at the  C in the B 
layer, resulting in the an enhanced $\sigma$- 
$\pi$ hybridization. This hybridization may get further accentuated in 
the Cu substituted samples due to the presence of 3d orbitals of Cu, leading to an 
increased interband scattering and consequently in the pronounced variation of 
$\rho_{0}$ with T$_{C}$ in these samples (cf. Fig. 6). 
\begin{figure}
\includegraphics*[width=7cm,height=7cm]{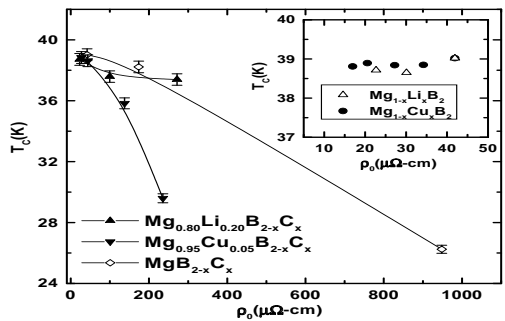}
\caption{\label{fig6}Plot of $\rho_{0}$ versus T$_{C}$ where $\rho_{0}$ is the residual resistivity obtained from 
the Bloch-Gruneisen fit of the normal state resistivity data, in all the series 
studied. Inset shows the data in the Cu and Li substituted samples. The solid 
lines are a guide to the eye.} 
\end{figure}

\subsection{3.7 Slope of high temperature resistivity} 
 In order
to obtain a quantitative measure of the temperature dependence
of resistivity in the various samples,  the magnitude of
d$\rho$/dT from the linear regime of $\rho$(T)  can be extracted.
In Fig.7a and Fig.7b are shown the 
$\rho$(T) data in the 200K-300K temperature range for two representative
series viz., Mg$_{1-x}$Li$_{x}$B$_{2}$ and MgB$_{2-x}$C$_{x}$. It
is clear from the 
Figure that in the Li substituted samples the slope decreases,
while in the C substituted samples the d$\rho$/dT increases with
substitution.
\begin{figure}
\includegraphics*[width=8cm,height=7cm]{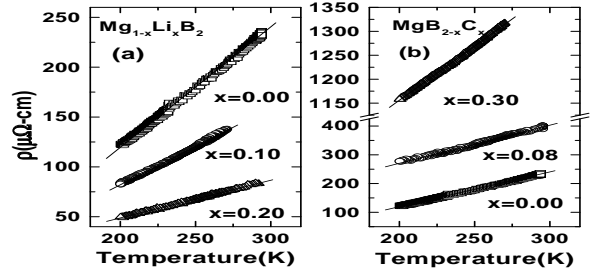}
\caption{\label{fig7}$\rho$(T) in the 200-300K range (a)in
Mg$_{1-x}$Li$_{x}$B$_{2}$ and in (b) MgB$_{2-x}$C$_{x}$, the
solid lines are the linear fits to the data.} 
\end{figure}
Similarly the slopes are obtained 
from a linear 
fit of the $\rho$(T) data in all the samples and are tabulated in
Table.2, and also shown  
in two panels for the cation substituted samples and carbon
substitutions  in Fig.8a and Fig.8b respectively. A feature that clearly 
emerges from the data is that the slope decreases by a large extent due to Mg 
substitution, implying the temperature dependence
of $\rho$(T) decreases  
in these samples with substitution. In contrast, in the carbon substituted 
samples i.e., in MgB$_{2-x}$C$_{x}$ and in
Mg$_{0.80}$Li$_{0.20}$B$_{2-x}$C$_{x}$ the slope increases with
the  level of C substituted (cf. Fig. 8b), indicating that the temperature
dependence of $\rho$(T)  
increases with substitution. The degree of increase in 
d$\rho$/dT is also high in 
both the series. In the Mg$_{0.95}$Cu$_{0.05}$B$_{2-x}$C$_{x}$ 
series, however, d$\rho$/dT shows a small 
decrease. Plotted in Fig.9a and Fig.9b in two panels is the 
variation of $\rho_{0}$ in the 
different series with substitution. The variation in 
d$\rho$/dT and $\rho_{0}$ with the 
concentration of the substitutent are very similar excepting in the 
Mg$_{0.95}$Cu$_{0.05}$B$_{2-x}$C$_{x}$ series. 
\begin{figure}
\includegraphics*[width=8cm,height=7cm]{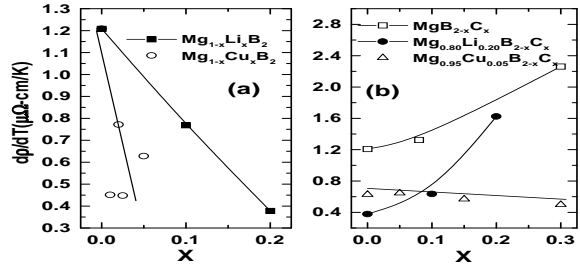}
\caption{\label{fig8} A plot of d$\rho$/dT versus the fraction substituted,
for (a)cation substituted samples and (b) for carbon substituted
samples.  The solid lines are a guide to the eye.} 
\end{figure}

\begin{figure}
\includegraphics*[width=8cm,height=7cm]{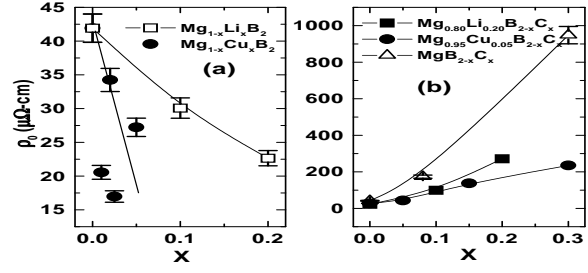}
\caption{\label{fig9}A plot of $\rho_{0}$ versus the substituted fraction in the
different series, (a) cation substituted samples (b) carbon
substituted samples. The solid 
lines are a guide to the eye. } 
\end{figure}
In order to obtain a qualitative understanding of the variation
in $\rho_{0}$ and d$\rho$/dT in the various series we take
recourse to an analysis of the conductivity in terms of the two band
model thought to be most appropriate to understand normal state
transport in MgB$_2$.\cite{mazin}
The expression for conductivity in a two band model\cite{mazin}
is given by 
\begin{equation}
\label{Eq. 5}
\frac{1}{\rho(T)} = \frac{1}{4\pi}\sum_{n=\sigma,\pi} \frac
{\omega_{pl,n}^{2}}{W_{n}(0,T)},
\end{equation}
where $\omega_{pl,n}$ is the plasma frequency for the band 'n'
and W$_{n}$(0,T) for n=$\sigma$ band is given by 
\begin{equation}
 W_{\sigma}(0,T)= \gamma_{\sigma} + \frac{\pi}{T}\int_{0}^{\infty}
 d\omega \frac {\omega}{sinh^{2}(\frac{\omega}{2T})} 
[\alpha_{tr}^{2}(\omega)F_{\sigma\sigma}(\omega)  +\alpha_{tr}^{2}(\omega)F_{\sigma\pi}(\omega)]
\end{equation}
where $\gamma_{\sigma} = \gamma_{\sigma\sigma}+\gamma_{\sigma\pi},$ and $\gamma_{nn'}$ is
the transition probability for an electron to scatter from band
index n to n' and $\alpha_{tr}^{2}(\omega)F_{nn'}(\omega)$ are the
transport Eliashberg functions. The expression W$_{\pi}$(0,T) can
be obtained by substituting $\pi$ for $\sigma$ in Eq. (6). 
It is clear from the Eq. (5)  that if the $\sigma$ band
i.e; if the first term in Eq. (5) dominates, the
temperature dependence of conductivity will be large, since 
this band is known to couple strongly with phonons in MgB$_{2}$.
On the other 
hand if the $\pi$ band dominates conduction, the temperature
dependence of
conductivity will be small as the $\pi$ band couples less effectively
with phonons in this system.
The reduced d$\rho$/dT in the cation substituted samples(cf. Fig.8a) and the
enhanced d$\rho$/dT in the C substituted samples (cf. Fig. 8b)
therefore imply that the $\pi$ band 
dominates conduction in the former
and the $\sigma$ band dominates conduction in the latter.

\vskip 2mm
\noindent
The domination of the $\sigma$ band in the carbon substituted
samples (MgB$_{2-x}$C$_{x}$ and
Mg$_{0.80}$Li$_{0.20}$B$_{2-x}$C$_{x}$)  would be possible if either
the plasma frequency for that band is large or if the
$\gamma_{\pi\pi}$ is large in Eq.5. Since it has been inferred from 
band structure calculations that the
$\omega_{pl,\sigma}$\cite{mazin2,mazin} is small in 
MgB$_{2}$, it appears that $\gamma_{\pi\pi}$ is large in the C
substituted samples. This increase in the $\pi-\pi$ scattering, 
could arise due to the presence of disorder along the 
C-axis as a result of the proximity of the p$_{z}$ electrons of
carbon to 
this region.  The increase in the d$\rho$/dT with increase in C
content would naturally  
follow due to the progressive increase in $\gamma_{\pi\pi}$  
consequent to an increase in the 
number of such scattering centres. 
The increase in $\rho_{0}$ with C 
substitution which is clearly seen from panels showing the C
substitutions in  Fig. 9b could arise due to increase in
$\gamma_{\sigma\pi}$, which is also apparent from Fig.6. In the
Mg$_{0.95}$Cu$_{0.05}$B$_{2-x}$C$_{x}$ series  
the interband scattering term $\gamma_{\sigma\pi}$ is very large
(cf. Fig.6).
This could make both the $\sigma$ and  
$\pi$ channels of 
conduction in Eq.5 equally probable, because of which drawing
inferences about the behaviour of either $\rho_{0}$ or
d$\rho$/dT becomes difficult.
\noindent 
\vskip 1mm 
The behaviour of d$\rho$/dT in the cation substituted samples
shows that despite substitution in Mg sublattice
the observed temperature dependence of $\rho(T)$ is 
smaller than in MgB$_{2}$, contrary to the findings of the theoretical
calculations\cite{mazin} which indicate that the $\gamma_{\pi\pi}$  
would be large, leading to the domination of the conductivity by
the $\sigma$ band.
However the larger contribution to conductivity from the $\pi$ band
(term2 Eq.5), observed in the cation substituted samples (cf.
Fig. 8a) could occur if   
the relative magnitudes of the plasma frequencies from the
$\sigma$ and $\pi$ bands get strongly affected due 
to these substitutions. This is highly plausible as it has been
demonstrated from band structure calculations of
LiB$_{2}$\cite{mazin2} that 
the band disposition and dispersion are strongly altered with
respect to that in MgB$_{2}$.  It can be seen 
from Fig. 9a that $\rho_{0}$ decreases with substitution in the 
cation substituted 
samples, a result rather surprising. These would imply that the
intraband $\pi-\pi$ scattering is progressively reduced due to these
substitutions. Band structure calculations would be necessary to
verify these 
experimental observations. Measurements  of the normal stste
resistivity in the Mg$_{1-x}$Al$_{x}$B$_{2}$ in which detailed
band structure calculations exists are in progress.
\section{4. SUMMARY and CONCLUSION} 
The variation of T$_{C}$ has been studied as a function of the 
variation in electron 
and/or hole concentration by appropriate chemical substitutions. 
A plot of T$_{C}$ 
versus the electron count estimated from the different series studied shows a 
universal behaviour, remaining flat for electron counts lesser
than the MgB$_{2}$  
value whereas it steeply decreases with electron count in excess of that in 
MgB$_{2}$. This has a striking similarity with the variation of 
hole DOS with energy. 
The ratio of the interband/intraband scattering seems to be 
strongly affected by 
the nature of the chemical dopant. The addition of C to the 
Boron layer  seems to increase    
$\gamma_{\sigma\pi}$ 
resulting in an increase in the residual resistivity and to the
depletion of T$_{C}$, which gets 
further enhanced  due to Cu substitution in the Mg sub-lattice. 
The slope of 
the resistivity curve in the 200K-300K, gives an indication of
the magnitude of  
the temperature dependence. The larger temperature dependence in the C 
substituted samples indicates that the $\sigma$-band dominates
conduction in this  
system. The temperature dependence of resistivity in the monovalent cation 
substituted samples are lowered with respect to that in
MgB$_{2}$, suggesting an  
enhanced participation in conductivity due to the $\pi$-bands. These results
clearly demonstrate that  
by selective chemical substitutions conductivity from certain bands can be 
probed.


\begin{thebibliography}{50} 
\bibitem{nagamatsu} 
{Jun Nagamatsu, Norimasa Nakagawa, Takahiro Muranaka, Yuji Zenitani, Jun 
Akimitsu , Nature 410, 63 (2001)} 
\bibitem{budko} 
{S.L.Bud'ko, G.Lapertot, C.Petrovic, C.E.Cunningham, N.Anderson and 
P.C.Canfield Phys. Rev. Lett. 86, 1877 (2001)} 
\bibitem{hinks} 
{ D.G.Hinks, H.Claus, J.D.Jorgensen, Nature 411, 457 (2001)} 
\bibitem{lorentz} 
{ B.Lorentz, R.L.Meng, C.W.Chu, cond-mat/0102264} 
\bibitem{tomita} 
{T.Tomita, J.J.Hamlin, J.S.Schilling, D.G.Hinks, J.D.Jorgensen, Phys. Rev. B 
64, 092505 (2001) } 
\bibitem{kortus} 
{ J.Kortus, I.I.Mazin, K.D.Belashchenko, V.P.Antropov, L.L.Boyer Phys. Rev. 
Lett. 86, 4656 (2001) } 
\bibitem{an} 
{ J.M.An, W.E.Pickett, Phys. Rev. Lett. 86, 4366 (2001) } 
\bibitem{yildrim} 
{ T.Yildirim, O.Gulseren, J.W.Lynn, C.M.Brown, T.J.Udovic, Q.Huang, N.Rogado, 
K.A.Regan, M.A.Hayward, J.S.Slusky, T.He, M.K.Haas, P.Khalifah, K.Inumaru, 
R.J.Cava, Phys. Rev. Lett. 87, 37001 (2001) } 
\bibitem{kong} 
{ Y.Kong, O.V.Dolgov, O.Jepsen, O.K.Andersen, Phys. Rev. B 64, 020501R (2001) } 
\bibitem{liu} 
{ Amy.Y.Liu, I.I.Mazin, Jens Kortus, Phys. Rev. Lett. 87, 087005 (2001) 
\bibitem{bouquet} 
F.Bouquet, R.A.Fisher, N.E.Philips, D.G.Hinks, J.D.Jorgensen, Phys. Rev. 
Lett. 87, 047001 (2001)} 
\bibitem{giubileo} 
{F.Giubileo, D.Roditchev, W.Sacks, R.Lamy, D.X.Thanh, J.Klein, S.Miraglia, 
D.Fruchart, J.Marcus, Ph.Monod, Phys. Rev. Lett. 87, 177008 (2001) } 
\bibitem{iavorone}
{M. Iavarone, G. Karapetrov, A.E. Koshelev, W. K. Kwok, G. W.
Crabtree and D. G. Hinks, Phys. Rev. Lett. {\bf 89} 187002 (2002)} 
\bibitem{schmidt}
{H. Schmidt,  J. F. Zasadzinski, K. E. Gray and D. G. Hinks,
Phys. Rev. Lett.  {\bf 88}, 127002 (2001)}
\bibitem{chen}
{X. H. Chen, M. J. Konstantinovic, J. C. Irwin, D. D. Lawrie and
J. P. Franck, Phys. Rev. Lett. {\bf 87} 157002 (2001)}
\bibitem{slusky} 
{ J.S.Slusky, N.Rogado, K.A.Regan, M.A.Hayward, P.Khalifah, T.He, K.Inumani, 
S.M.Loureiro, M.K.Haas, H.W.Zanderbergen, R.J.Cava , Nature 410, 343 (2001) } 
\bibitem{postorino} 
{P.Postorino, A. Congeduti, P. Dore, A. Nucara, A.Bianconi, D.
Di Castro, S. De Negri and A. Saccone, Phys. Rev. B {\bf 65}
020507(R) (2001) } 
\bibitem{lorentz1} 
{B.Lorentz, R.L.Meng, Y.Y.Xue, C.W.Chu, Phys. Rev. B 64, 052513 (2001) } 
\bibitem{cimberele} 
{ M.R.Cimberle, M.Novak, P.Manfrinetti and A.Palenzona, Supercond. Sci. 
Technol. 15, 43-47 (2002) } 
\bibitem{zhao} 
{Y.G.Zhao, X.P.Zhang, P.T.Qiao, H.T.Zhang, S.L.Jia, B.S.Cao, M.H.Zhu, 
Z.H.Han, X.L.Wang, B.L.Gu, Physica C 361, 91-94 (2001)} 
\bibitem{syli} 
{ S.Y.Li, Y.M.Xiong, W.Q.Mo, R.Fan, C.H.Wang, X.G.Luo, Z.Sun, X.T.Zhang, L.Li, 
L.Z.Cao, X.H.Chen, Physica C 363, 219-223 (2001) } 
\bibitem{toulemonde} 
{ P.Toulemonde, N.Musolino, H.L.Suo, R.Flukiger, cond-mat/0207033 
Superconductivity Science and Technology 17-19, (2002)} 
\bibitem{moritomo} 
{Y.Moritomo, Sh.Xu , cond-mat/0104568} 
\bibitem{kazakov} 
{ S.M.Kazakov, M.Angst, J.Karpinski, I.M.Fita, R.Puzniak, cond-mat/0103350} 
\bibitem{xu} 
{Sheng Xu, Yutaka Moritomo, Kenichi Kato, Arao Nakamura, cond-mat/0104534} 
\bibitem{gasporov} 
{ Vitaly A.Gasparov, N.S.Sidorov, I.I.Zver'kova, M.P.Kulakov, cond-mat/0104323 
JETP Letters, April 12 (2001)} 
\bibitem{kalavathi} 
{ S.Kalavathi, A.Bharathi, S.Jemima Balaselvi, G.L.N.Reddy, V.S.Sastry, 
Y.Hariharan, T.S.Radhakrishnan, Solid State Physics (India) 44 (2001)} 
\bibitem{pena}
{ O. dela Pena, A. Aguayo and R. des Coss Phys. Rev. B {\bf 66}
012511 (2002)} 
\bibitem{ahn1} 
{J.S.Ahn, Young-Jin Kim, M.-S.Kim, S.-I.Lee, E.J.Choi, cond-mat/0202415} 
\bibitem{ahn2} 
{ J.S.Ahn, E.S.Choi, W.Kang, D.J.Singh, E.J.Choi, cond-mat/0202457} 
\bibitem{zhang} 
{ Shao-ying Zhang, Jian Zhang, Tong-yum Zhao, Chuan-bing Rong, Bao-gen Shen, 
Zhao-hua Cheng, cond-mat/0103203} 
\bibitem{paranthaman} 
M.Paranthaman, J.R.Thompson, D.K.Christen, cond-mat/0104086 
\bibitem{jsahn} 
{Jai Seok Ahn, Eun Jip Choi, cond-mat/0103169} 
\bibitem{mickelson} 
{W.Mickelson, John Cumings, W.Q.Han, A.Zettl, Phys. Rev. B 65, 052505 (2002) } 
\bibitem{takenobou} 
{ T.Takenobu, T.Ito, Dam.H.Chi, K.Prassides, Y.Iwasa, cond-mat/0103241} 
\bibitem{bharathi} 
{ A.Bharathi, S.Jemima Balaselvi, S.Kalavathi, G.L.N.Reddy, V.Sankara Sastry, 
Y.Hariharan, T.S.Radhakrishnan, Physica C 370, 211-218 (2002) } 
\bibitem{mehl} 
{ M.J.Mehl, D.A.Papaconstantopoulos, D.J.Singh, Phys. Rev. B 64, 140509R 
(2001) } 
\bibitem{rosner} 
{ H.Rosner, A.Kitaigorodsky, W.E.Pickett, Phys. Rev. Lett. 88, 127001 (2002) } 

\bibitem{finnemore} 
{D.K.Finnemore, J.E.Ostenson, S.L.Bud'ko, G.Lapertot, P.C.Canfield, Phys. 
Rev. Lett. 86, 2420 (2001)} 
\bibitem{canfield} 
{P.C.Canfield, D.K.Finnemore, S.L.Bud'ko, J.E.Ostenson, G.Lapertot, 
C.E.Cunningham, C.Petrovic, Phys. Rev. Lett. 86, 2423 (2001)} 
\bibitem{putti} 
{ M.Putti, E.Galleani d'Agliano, D.Marre.F.Napoli, M.Tassisto, P.Manfrinetti, 
A.Palenzona, C.Rizzato, S.Massidda, Eur. Phys. J. B, 25, 439-443 (2002) } 
\bibitem{mazin2}
{I. I. Mazin and V. P. Antropov, Physica C {\bf 385}, 49 (2003)}
\bibitem{mazin} 
{I.I.Mazin, O.K.Andersen, O.Jepsen, O.V.Dolgov, J.Kortus, A.A.Golubov, 
A.B.Kuz'menko, D.van der Marel, Phys. Rev. Lett. 89, 107002 (2002)} 
\bibitem{golubov}
{A.A. Golubov and I. I. Mazin, Phys. Rev. B {\bf 55} 15146 (1997)}
\end{thebibliography}
\end{document}